# A Low Cross-Polarization Slotted Ridged SIW Array Antenna Design With Mutual Coupling Considerations

Alireza Mallahzadeh, *Senior Member, IEEE*, and Sajad Mohammad-Ali-Nezhad, *Member, IEEE*

*Abstract*—A novel shunt longitudinal slot planar array antenna based on a ridged substrate integrated waveguide (RSIW) with a good cross polarization and isolation is designed and fabricated. The slots in the proposed structure are placed on the centerline of the broad wall of the substrate integrated waveguide (SIW), in order to reduce cross polarization. Also, the asymmetric field distributions on both sides of each slot, which are required for making radiation possible, are realized through using ridges close to the narrow walls of the SIW. The admittance of a single slot in the proposed structure is obtained regarding the different values of the main parameters of the structure. An eight-element longitudinal slot array antenna on the broad wall of the RSIW is designed based on Elliot's design procedure. In this design, internal and external mutual coupling effects are taken into account through calculating the active admittance for each slot of the array. A 2-D graph is presented which contains the real and imaginary parts of the equivalent admittance of a slot based on the changes in the widths of the ridges and length of the slot. Using this graph, the widths of the ridges and length of the slot can be revised in a way that the mutual coupling effect is taken into account. Since the array antenna is composed of collinear elements, it demonstrates a better cross polarization and isolation in comparison with conventional slot array antennas. Simulation results show good agreement with measurement results of the fabricated proposed antenna.

## I. INTRODUCTION

SINCE longitudinal shunt slotted waveguide array antennas have low loss, high gain, high efficiency, simple feeding, and an easy manufacturing process, they have a wide range of applications in radar and communication systems [1]–[7]. However, this structure has some drawbacks.

In order to have radiating slots, they are required to have some distance from the centerline of the substrate integrated waveguide (SIW) broad wall; also for having an in-phase radiation from an array of slots, the slots have to be displaced on alternate sides of the centerline of the broad wall; both of these situations result in an increase in the cross polarization of the structure and contribute to the appearance of second-order beams, which degrades the antenna efficiency [8], [9].

There are different methods for reducing cross polarization and removing second-order beams such as using corrugated narrow walls [10], exciting untilted slots on the narrow wall using a dielectric plate with conducting strips on both sides [11], using tilted wires in slotted waveguides [12] or using irises close to the slots [13]. However, manufacturing the mentioned structures is often difficult and obtaining a proper side lobe level (SLL) in these structures is not easy. Another method is using ridges inside a waveguide, such as the wiggly [14], asymmetric [15], and tilted [16] single-ridged waveguide, and the concave and convex double-ridged waveguide [17], [18] slot array antennas.

Another drawback of the longitudinal slotted waveguide array antennas is the width of the waveguides which is $0.7\lambda_0$, where $\lambda_0$ is the free-space wavelength. In planar arrays, this waveguide width limits the E-plane scanning angle to $\pm 25°$ with respect to the array antenna boresight. For having a scanning angle up to the endfire position, without producing grating lobes, waveguides with a width of $0.5\lambda_0$ are needed. This $0.5\lambda_0$ waveguide width can be realized using ridges inside the waveguides [19]–[21].

In recent years, substrate-integrated waveguides have received much attention in microwave and millimeter wave applications [22]–[24]. Slotted SIW array antennas are used in different systems to make use of the advantages of the SIW technology over conventional waveguides, such as having low cost and simple manufacturing process and also being able to easily integrate with planar circuits [25]–[31].

Due to the presence of dielectrics in slotted SIW antennas, the width of SIWs is narrower compared to that of the conventional waveguides. In these structures, the width of SIWs is about $0.5\lambda_0$. However, due to the low height of SIW structures and presence of dielectric material inside them, mutual coupling between array elements is increased which adversely affects the antenna scanning range and impedance bandwidth. On the other hand, due to displacement of the slots on both sides of the SIW centerline, cross polarization is increased.

In this paper, a two-layer collinear slot array antenna based on **ridged substrate integrated waveguides** (RSIWs) with a cross polarization of about $-40$ dB and a radiation pattern with an SLL of $-25$ dB is presented. Using ridges in the proposed SIW structure, the electric field distribution is controlled in a way that radiation is made possible for a slot placed on the centerline of the SIW broad wall. In Section III, an array design procedure is proposed. The changes in the antenna admittance versus the changes in the ridge and slot dimensions are investigated. Also, external and internal mutual coupling effects are taken into account in Elliot's array design procedure as an active admittance calculated using CST software and the mutual admittance for each slot is calculated. A two-dimensional (2-D) graph is presented so that the dimensions of the ridges and the slots can be revised while considering the active admittances. In Section IV, a planar array antenna is designed and manufactured. Due to the low cross polarization and high isolation between linear arrays of the proposed structure, the scanning range could be improved comparing to that of the conventional slot array antennas. Simulation and measurement results are compared and show good agreement.

## II. SINGLE-SLOTTED RSIW ANTENNA OPERATION PRINCIPLES

In order to have a planar array with an appropriate scanning range in the E-plane, the linear arrays should have a proper width. SIW structures can provide designers with a proper width of less than $0.5\lambda_0$, but there is a high mutual coupling between the radiating elements of the SIW structure due to its low height that results in a poor isolation between the linear arrays. It has to be mentioned that mutual coupling is one of the main factors that negatively affects array performance and results in a distorted array radiation pattern. Scan blindness is also caused by poor mutual coupling [33], [34]. Moreover, obtaining the desired radiation pattern requires a proper arrangement of the radiating elements in a way that results in a structure with a low cross polarization and no second-order beams. The proposed structure can be used to realize these goals.

The distance between the slots and SIW centerline results in an increase in the cross polarization of array antennas; for eliminating this problem, the slots must be placed on the centerline of the structure. On the other hand, only those slots can radiate such that the field distribution on both sides of them inside the SIW is asymmetric, i.e., the slots cut the current distribution lines on the upper plane of the SIW. When SIW structures operate in the dominant mode, the field distribution is symmetrical regarding the centerline of the SIW and the slot cannot cut the electric current distribution lines.

For having radiating slots with no distance from the centerline of the SIW, the field distribution on both sides of each slot can be made asymmetric, instead of having distanced slots from the centerline with a symmetrical field distribution inside the SIW structure. For this purpose, two ridges on both sides of the slot can be used, as shown in Fig. 1. Similar to [32], however in that paper, a long slot leaky-wave antenna is presented.

These two ridges with different dimensions create an asymmetric field distribution on both sides of the slot. The top and middle planes of the two-layer SIW structure are shown in Fig. 1. The width of the lower layer of the SIW is less than that of the upper layer; in fact, the lower layer originally has the same width as of the upper layer, but the two ridges on the lower layer make it narrower. It has to be mentioned that placing the vias close to the two longitudinal edges of the lower layer is not necessary.

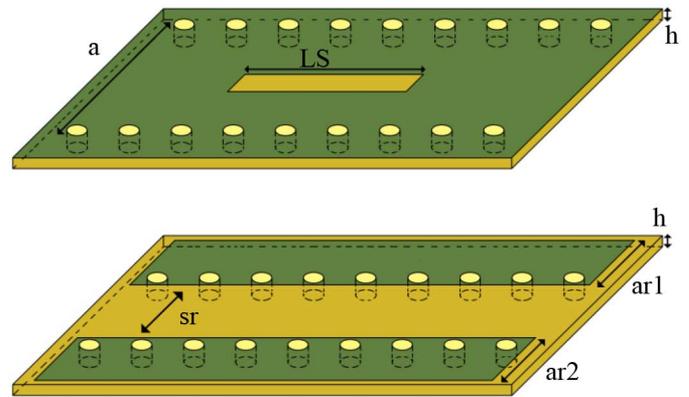

Fig. 1. Top and bottom layers of the proposed slotted RSIW antenna.

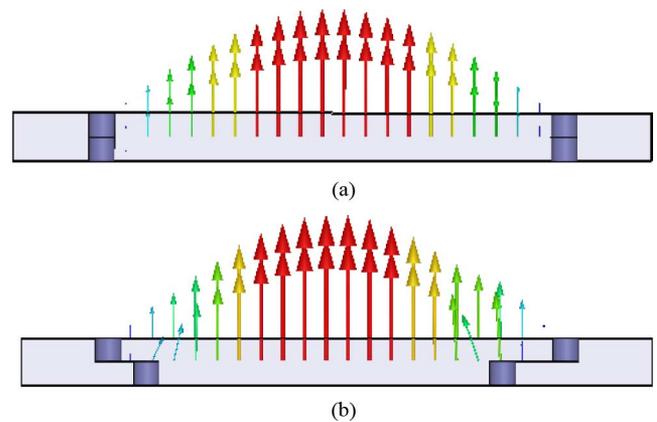

Fig. 2. Electric field in cross section of (a) SIW and (b) RSIW.

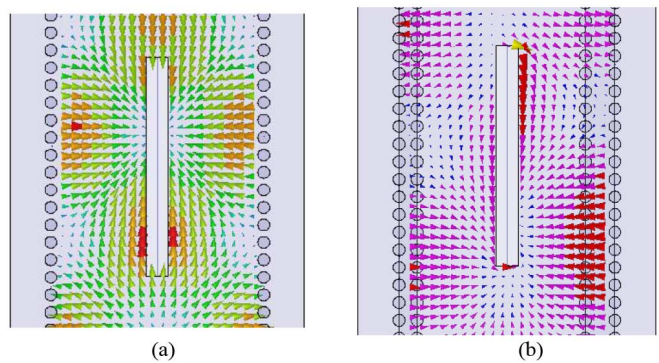

Fig. 3. Current distribution on broad wall of (a) SIW and (b) RSIW.

The electric current and field distribution of SIW and double-ridged SIW structures are compared in Figs. 2 and 3. Electric fields in the cross-sectional view of the SIW and RSIW structures, with no slots, are shown in Fig. 2. As shown in Fig. 2, the electric field in the SIW is completely symmetrical regarding the centerline of the structure, whereas in the RSIW structure, with two asymmetric ridges with different widths placed close to the SIW edges, the electric field is not symmetrical. In the slotted RSIW structure, the maximum intensity of the electric field does not occur in the center of the structure. In fact, the ridge placed on the right side, which is wider than the ridge on the left side, causes the electric field maximum intensity to move toward the left side of the structure.

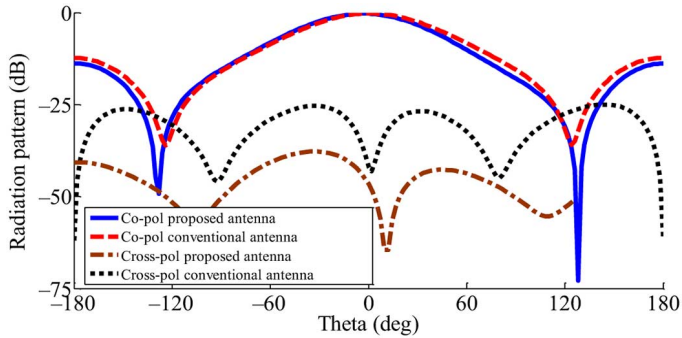

Fig. 4. Co- and cross-polarization E-plane radiation patterns of the proposed and conventional antennas.

Fig. 3 shows the electric current distribution lines around a slot located on the centerline of the broad wall of SIW and RSIW structures. In the SIW structure, the electric current distribution lines do not cut the slot, but in the RSIW structure, the slot is cut by the current distribution lines and the electric current flows around the slot that results in the formation of an electric field inside the slot and consequently radiation by a slot located on the centerline.

E-plane radiation patterns for the conventional and proposed slotted SIW antennas are presented in Fig. 4. As shown in this figure, cross polarization for the proposed slotted RSIW is about 10 dB better than that of the slotted SIWs with distanced slots from the centerline. This improvement is due to placing the slots on the centerline of the proposed RSIW structure.

Radiation characteristics of the slotted RSIW antennas can be altered by changing the dimensions of the ridges. Values of radiation admittances and resonance lengths of the slots can be obtained for different dimensions of the ridges. For having radiating slots and also realizing RSIW slots with different admittance values which are suitable for being used as the radiating elements of an array with low SLL, two ridges with different dimensions from each other are needed. In this way, asymmetric field distributions in the slot positions can be generated. In order to avoid a bulky structure and also increase in the number of layers of the structure, the height of the ridges can be kept fixed and their widths can be changed with respect to each other. Also for having a fixed propagation constant and operation frequency in the proposed structure, the distance between the ridges must be fixed, i.e., the width of the lower SIW layer that contains the two ridges has to be fixed (sr is fixed) and only the widths of the ridges (ar1 and ar2) are changed with respect to each other, so that the asymmetric field distribution around the centerline is realized.

Fig. 5 shows changes in the propagation constant versus frequency as a result of changes in the sr for ar1 = 0.5 mm and ar1 = 0 mm (note that ar1 = 0 makes the structure single ridged).

The data presented in Fig. 5 are obtained based on a proper SIW structure that can operate in the dominant mode at the frequency of 10 GHz. The initial dimensions of the structure are d = 1.2 mm, p = 0.8 mm, h = 0.8 mm, and a = 13 mm with an $\varepsilon_r$ of 3.3, where d and p stand for the distance between the vias and the diameter of the vias, respectively.

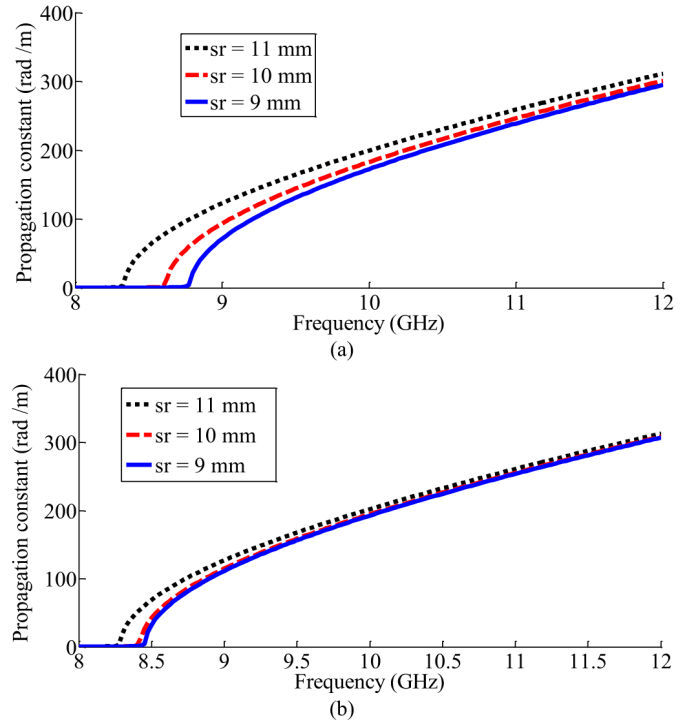

Fig. 5. Propagation constant of various widths of the lower layer of the RSIW (sr) for (a) ar1 = 0.5 mm and (b) ar1 = 0 mm (single ridged).

Changing sr causes a change in the propagation constant and consequently changes the guided wavelength of the structure ($\lambda_g$). Since the distance between the slots in the array design equals $\lambda_g/2$ and the distance between the last slot and the short circuited end of the structure is $\lambda_g/4$, having a fixed $\lambda_g$ which depends on having a fixed sr is of vital importance. Although the asymmetric field distribution in the position of the slots can be realized through using only a single ridge close to one of the SIW structure edges, using a single ridge with a variable width results in a variable propagation constant which is not appropriate for designing structures with a desired SLL. The reason for changing the widths of the ridges is that a nonuniform field distribution in the structure and different amounts of power radiated from the slot is needed for achieving a desired SLL.

When the structure has a single ridge, i.e., the second ridge has zero width, sr equals the distance between the first ridge and the narrow wall on the opposite side of the SIW (sr = a − ar1). Since the width of the SIW is fixed, changes in the width of the ridge directly affect the value of sr. Therefore, when the width of the ridge is changed for tapering the field distribution at the antenna aperture, sr is changed as well. But in the situation in which two ridges exist in the SIW and sr = a − ar1 − ar2, both ar1 and ar2 can be changed simultaneously in a way that sr is kept fixed, and $\beta$ and $\lambda g$ are kept fixed as a result. This is while in the single-ridged SIW, sr is inevitably changed when ar1 is altered. It is also clear that $\beta$ changes according to the changes in sr and results in defects in the tapering mechanism of the antenna. Based on the previous explanations, it is concluded that the single-ridged SIW cannot be used and in double-ridged SIWs, both ar1 and ar2 must be altered in a way that sr is kept fixed.

sr has to be selected in way that higher order modes are not generated and changing the ratio between ar1 and ar2, for obtaining different radiated amounts of power from the slots, is made possible. In the proposed structure, sr = 11 mm.

## III. Design Principles of the Proposed Array Antenna

The array antenna can be designed based on the proposed array element by going through the four steps described as follows.

Step 1) In order to obtain an appropriate SLL, the slots should be fed in a nonuniform manner. For this purpose, the equivalent admittance of a single slot can be found for different widths of the ridges and lengths of the slot using a full-wave EM software. The length of the slot should be selected in a way that its corresponding admittance has zero imaginary part, i.e., the resonance slot length is obtained. Different widths of the ridges result in different conductances which can be used for tapering voltage or current distributions of array elements.

Step 2) For achieving the desired SLL, the coefficients for exciting the slots can be found using Chebyshev amplitude distribution. Different admittance values needed for realizing Chebyshev coefficients can be obtained through choosing proper widths for the ridges and lengths for the slots. Since mutual coupling is not considered in the design of the proposed structure, the obtained radiation pattern does not fully comply with the desired radiation pattern. Input matching can also be improved through considering the mutual coupling effects.

Step 3) A 2-D graph is produced based on the results obtained in Step 1. In this graph, conductance and susceptance values with their corresponding ar1 and LS parameters values are presented. ar1 and LS are shown on the horizontal and vertical axes of the graph, respectively.

Step 4) For each of the slots of the array antenna designed in Step 2, the active admittance is calculated regarding the internal and external mutual couplings. Active admittance values must be considered in the array design. For the sake of simplicity in the design procedure and also due to the low values of mutual coupling in the proposed antenna, mutual admittance values are added to the self-admittance values, resulting in slots with complex admittance values. For having resonance admittances, i.e., admittances with zero imaginary part, the length of each slot should be chosen in a way that the imaginary part of its equivalent admittance equals the imaginary part (susceptance) of its corresponding mutual admittance, but with an opposite sign.

Conductance values of the mutual admittances are added to the conductance values of the slots calculated without considering the mutual coupling in Step 2. In order to obtain the desired SLL, the ratio between the slot conductances must comply with the Chebyshev coefficients found in Step 2. But when the conductance values are changed, due to adding the mutual admittances, they do not match with the Chebyshev coefficients found in Step 2 anymore. Therefore, the initial conductance values found in Step 2 must be changed according to the corresponding conductance of mutual admittance for each slot so that when the mutual coupling effect is included, the conductance values become equal to the initial conductance values of Step 2 and match with the desired Chebyshev coefficient. For this purpose, after calculating slots' lengths in the previous step with mutual coupling effect included, the values of new conductances and their corresponding widths of the ridges are obtained using 2-D proposed graph.

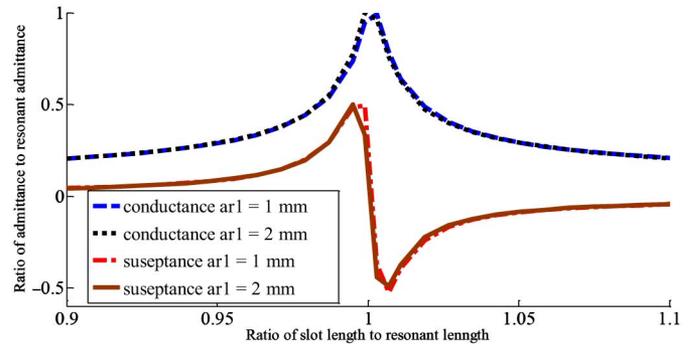

Fig. 6. Slot conductance and susceptance normalized to resonance conductance versus normalized slot length at 10 GHz.

For finding slots' lengths and widths of the ridges regarding the new admittance values, the 2-D graph produced in Step 3 is used. Finally, the new admittance values for each slot are obtained using

$$Y_n = Y_{nn} - Y_n^b = Y_{nn} - \left(G_n^b + jB_n^b\right) \quad (1)$$

where $Y_n^b$ is the mode–voltage–weight sum of the mutual admittances for $n$th slot.

### A. Admittance Properties of the Proposed Slotted RSIW Antenna

The ABCD matrix of the structure is obtained using CST software, which is similar to the ABCD matrix of shunt admittance. Therefore, the slotted RSIW antenna is equal to shunt admittance.

Since the admittance characteristics of the proposed antenna are an essential requirement in the slotted RSIW array antenna design procedure, the effects of the changes in the widths of the ridges and slot length on these characteristics have to be investigated.

Fig. 6 presents the ratio between the slot conductance and susceptance and the slot resonance conductance and susceptance versus the ratio between the slot length and the slot resonance length. The resonance of the slots occurs for lengths at which the imaginary parts of the impedances are equal to zero and the real parts have their maximum values.

For designing the nonuniform array antenna, the resonance slots' lengths and the resonance conductance for different widths of the ridges were calculated. Fig. 7 shows the normalized resonance length of a single slot versus the changes in ar2 while ar1 is fixed. It has to be mentioned that by increasing the asymmetry of the structure, i.e., increasing the ridge width, the width of the resonance slot is increased.

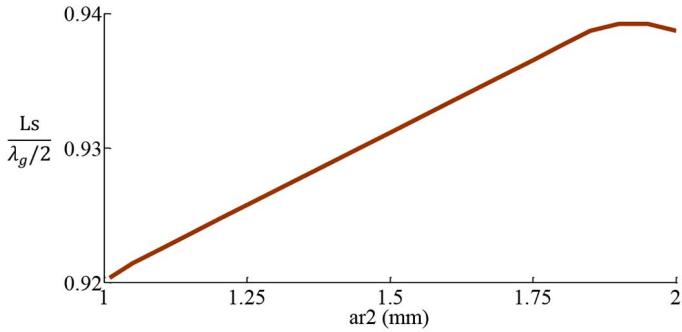

Fig. 7. Slot resonant length normalized to $\lambda_g/2$ versus ridge width at 10 GHz (sr = 11 mm).

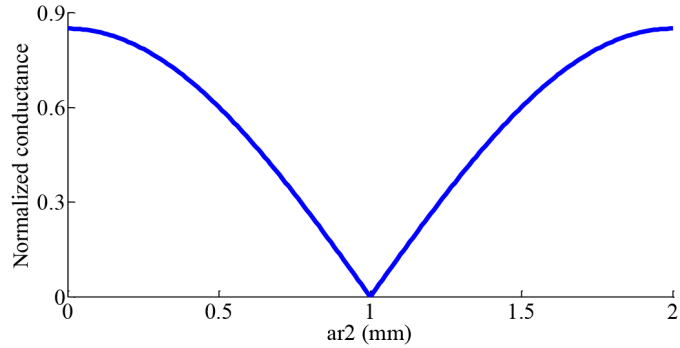

Fig. 8. Normalized resonant conductance versus ridge width at 10 GHz (sr = 11 mm).

Fig. 8 shows the resonant conductance normalized to the conductance of the RSIW versus the changes in ar2 while ar1 is fixed. As shown in Fig. 8, there is no radiation from the slots when ar1 = ar2, because in this situation, the structure is symmetrical regarding the position of the slots.

### B. Slotted RSIW Array Antenna Design Without Considering the Mutual Coupling

The top view of the upper layer (top plane) and the top view of the lower layer (middle plane) of the SIW structure are shown in Fig. 9. As shown in Fig. 9, the ridges close to the SIW edges are included in the bottom layer and the slots located on the centerline are placed on the top plane.

The center-to-center distance between the slots is $\lambda_g/2$, which causes a $180°$ phase difference between the powers radiated from adjacent slots. There are two ridges with different widths on both sides of each slot. In order to have in-phase fed slots and consequently an in-phase radiation from them, the widths of the ridges on both sides of the slots must be alternatively changed, i.e., if the width of the ridge on the left side of the slot number x is bigger than the width of the ridge on the right side of the same slot, for the next slot (slot number x + 1), the bigger ridge is on the right side and the smaller one is on the left side. Finally, the lower SIW layer is created with a zigzag shape. Correct calculation of $\lambda_g$ is an essential task which is done using CST software.

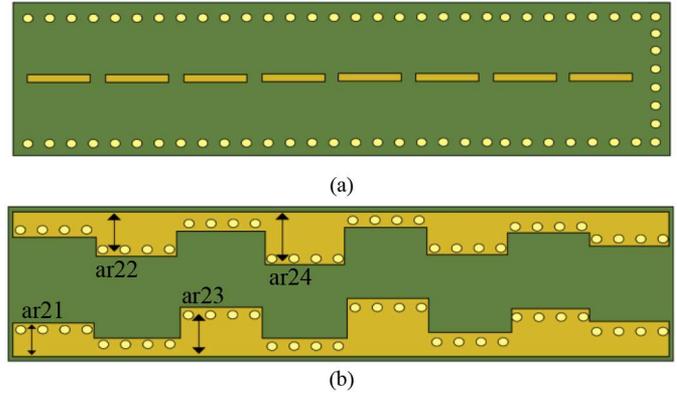

Fig. 9. Proposed collinear slotted RSIW array antenna (a) top plane (b) middle plane.

Since each slot is equal to a shunt admittance, for having a proper impedance matching, i.e., a termination at the end of the SIW structure with no effects on the equivalent circuit of the shunt admittances, the end of the proposed SIW structure is short circuited at a distance of $3\lambda_g/4$ from the center of the last slot and is considered as a zero admittance with no influence on the equivalent circuit of the proposed SIW structure.

The length of the slots should be selected in a way that resonance slots are obtained; in other words, the imaginary parts of slot admittances are equal to zero. The resonance length of the slots which is about $\lambda_0/2$ can be obtained with more accuracy, according to the widths of the slots, from Fig. 7.

In the pattern synthesis procedure, Chebyshev amplitude distribution can be used for exciting the slots to realize an SLL of $-25$ dB. For achieving this SLL, the admittance of each slot must be calculated based on Chebyshev coefficients. For having resonance slots with proper admittances, the widths of the ridges and their corresponding resonance slot lengths have to be taken into consideration. The admittance values proposed in Figs. 7 and 8 are calculated without considering the effect of other slots. However, the mutual coupling effect must be taken into account for calculating the admittance of the slots.

The lengths of the slots and the widths of the ridges are listed in Table I for an eight-element linear array, based on Chebyshev coefficients for an SLL of $-25$ dB. In the proposed structure, sr is fixed and is equal to 11 for having a fixed propagation constant.

These are the initial lengths of the slots and widths of the ridges in the linear array without taking into account the mutual coupling.

### C. 2-D Graph for Calculating the Admittance of the Proposed Slotted RSIW Antenna

In order to revise the admittances of the slots in a way that mutual coupling effects are taken into consideration, the widths of the ridges and lengths of the slots have to be varied.

TABLE I
SLOTS' LENGTHS AND WIDTHS OF THE RIDGES FOR THE
EIGHT-ELEMENT LINEAR ARRAY WITHOUT CONSIDERING
THE MUTUAL COUPLING EFFECTS

| Slot number | Ls | ar2 |
|---|---|---|
| 1 | 14.62 | 1.14 |
| 2 | 14.71 | 1.31 |
| 3 | 14.83 | 1.62 |
| 4 | 14.96 | 1.91 |
| 5 | 14.96 | 1.91 |
| 6 | 14.83 | 1.62 |
| 7 | 14.71 | 1.31 |
| 8 | 14.62 | 1.14 |

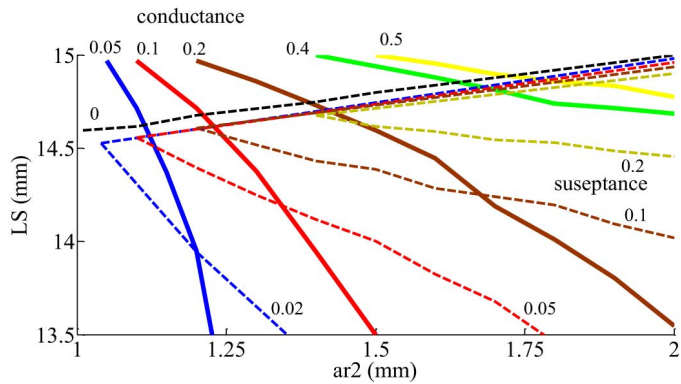

Fig. 10. Conductance and suseptance contour curve.

The new admittances have both the real and imaginary parts. For finding the new widths of the ridge and slot lengths regarding the new admittance values, the graph shown in Fig. 10 can be utilized. ar2 and LS values are shown on the vertical and horizontal axes of the mentioned graph, respectively. The solid lines show different values of the normalized conductances and the dashed lines show different values of the normalized susceptances. The desired admittance values are found at the points where these two sets of lines meet and then the widths of the ridges and lengths of the slots can be determined.

For obtaining Fig. 10, first conductance and susceptance values for a variable ar2 and fixed Ls were calculated using the full-wave CST software similar to Figs. 6–8. This is done for all values of Ls. In the next step, corresponding Ls and ar2 values for all given values of conductance or susceptance were obtained.

The widths of the ridge and slot lengths must be altered regarding the calculated mutual admittances. Since the new admittance values, which are combinations of the self-admittances and mutual couplings for each slot, have imaginary parts, proper lengths for the slots should be chosen in a way that susceptance values with an opposite singe of susceptances of the admittance values in the previous step are produced in each slot. By doing this, all of the resulted slots have real admittance values, but the resulted real admittance values do not comply with the Chebyshev distribution coefficients defined in the previous steps. Therefore, the initial real admittance values, as defined in Step 2 of Section III, should be revised in a way that the final real parts of the admittances, resulted from adding the active admittance to the self-admittance for each slot, match with the desired Chebyshev distribution coefficients calculated at the early stages of the design procedure.

### D. Slotted RSIW Array Antenna Design While Considering the Mutual Coupling Effect

In this paper, the width of each ridge and the length of each slot for a given aperture distribution are obtained through considering the internal and external mutual couplings; this method is similar to Elliot's method. In this method, admittance of each slot of the RSIW includes both self-admittance, as calculated in the previous section, and mutual admittance. Also in this paper, CST Microwave Studio package is used for calculating internal and external mutual couplings.

Equation 2 is used for finding the active admittance of the slots. For this purpose, all the slots except the nth slot have to be short circuited and by placing a current source at the mth slot and measuring the voltage at the nth slot, $Y_{mn}$ is calculated as

$$Y_{mn} = \frac{I_m}{V_n}. \quad (2)$$

When $Y_{mn}$ is known, Elliot's method can be used for designing the linear array. In the first step, the lengths of the slots and the widths of the ridges can be calculated based on Figs. 6–8. In these figures, self-admittance and the resonance lengths of the slots are presented without considering the mutual coupling effects.

Having the values of these assumed lengths, ar1 and ar2, $Y_{mn}$ can be calculated based on the described method. In the bottom of the next page, $Y_{mn}/Y_0$ matrix is presented (note that the values are multiplied by 100).

The mode–voltage–weight sum of the mutual admittances for each slot equals

$$Y_1^b = Y_8^b = 1.29 + i2.66$$
$$Y_2^b = Y_7^b = 2.62 + i2.99$$
$$Y_3^b = Y_6^b = 2.04 + i2.41$$
$$Y_4^b = Y_5^b = 2.24 + i7.49$$

To realize Chebyshev coefficients, the lengths of the slots and the widths of the ridges should be calculated again, based on the obtained mutual coupling admittances. The lengths of the slots should be selected in a way that the imaginary parts of the self-admittances are equal to the imaginary parts of the mutual couplings for each slot, but with an opposite sign. The widths of the ridges and the lengths of the slots in which the mutual coupling is taken into account are presented in Table II.

These results can be iterated, but as proved by Elliot, extra iterations do not change the results considerably; therefore, they can be acceptable.

The reflection coefficient of the collinear slotted RSIW linear array with and without mutual coupling effect versus frequency is presented in Fig. 11. Fig. 12 presents co- and cross polarization of H-plane radiation pattern with and without mutual coupling effect. Figs. 11 and 12 show that the antenna input matching and SLL are improved when internal and external mutual coupling effects are included in the calculations. Since the effect of other slots is included in the equivalent admittance of each slot and slots' lengths and widths of the ridges are revised, slots admittances do not contain an imaginary part and show a good matching with the input ports which have real impedances. The reason for improvement in the SLL is due to considering the effects of other slots in calculating the conductance of each slot resulting in more accurate equivalent conductance values and more precise calculation of Chebyshev distribution coefficients.

TABLE II
SLOTS LENGTHS AND WIDTHS OF THE RIDGES FOR THE EIGHT-ELEMENT LINEAR ARRAY WITH MUTUAL COUPLING EFFECT INCLUDED

| Slot number | Ls | ar1 |
| --- | --- | --- |
| 1 | 14.59 | 1.12 |
| 2 | 14.67 | 1.26 |
| 3 | 14.8 | 1.55 |
| 4 | 14.91 | 1.89 |
| 5 | 14.91 | 1.89 |
| 6 | 14.8 | 1.55 |
| 7 | 14.67 | 1.26 |
| 8 | 14.59 | 1.12 |

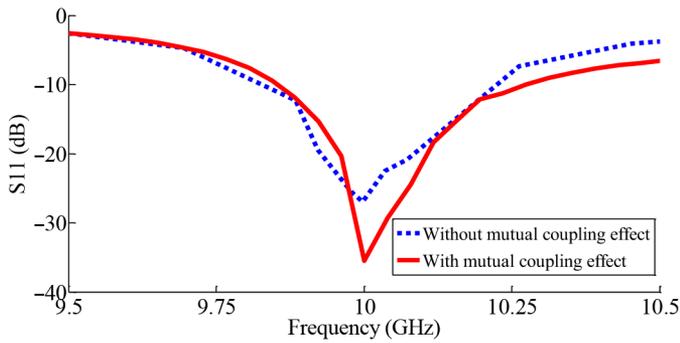

Fig. 11. Return loss of the proposed linear array antenna with and without mutual coupling.

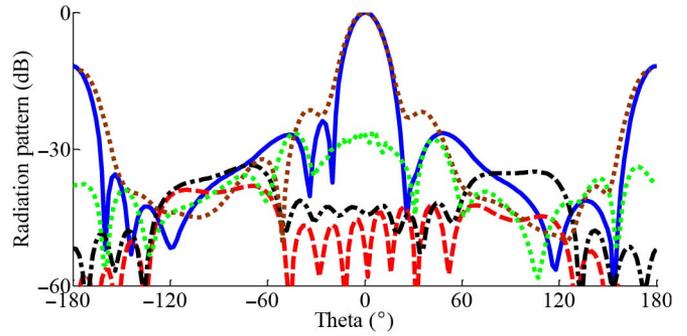

Fig. 12. Co- and cross-polarization radiation patterns of the proposed linear array antenna with and without mutual coupling effect and cross polarization of the conventional SIW linear array antenna at 10 GHz.
— Co-polarization with mutual coupling effect;
- - - cross polarization with mutual coupling effect;
⋯⋯ co-polarization without mutual coupling effect;
—·— cross polarization without mutual coupling effect;
⋯⋯ cross polarization of the conventional slotted SIW array antenna.

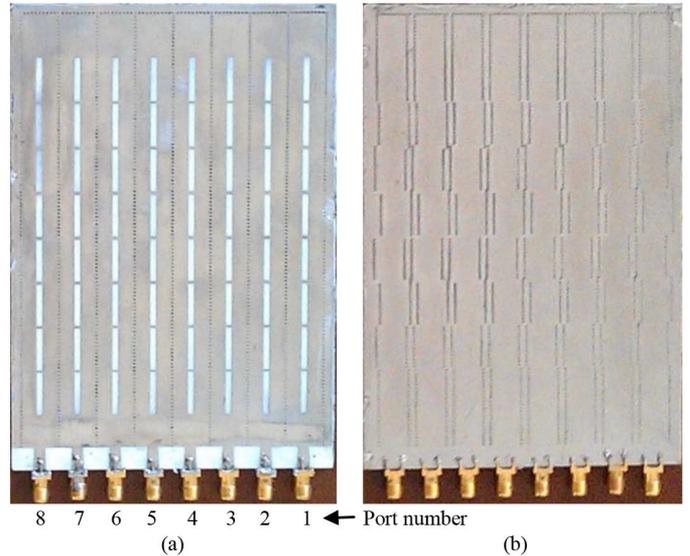

Fig. 13. Fabricated eight-element slotted RSIW linear array antenna. (a) Top plane. (b) Bottom plane.

Since the slots in the proposed structure are located on the centerline and are not alternatively placed on both sides of the centerline, cross polarization is improved in comparison with the conventional slotted SIW structures and is about $-43$ dB. Conventional slotted SIW array antennas have a cross polarization of about 27 dB; therefore, cross polarization of the proposed antenna has been improved for about 16 dB (Fig. 13).

Isolation in the planar array composed of the proposed linear array is lower than that of the conventional planar arrays. This is due to the low cross polarization of the proposed linear array. Eight uniform linear arrays are placed next to each other $(1 \times 8)$. In these linear arrays, ar1 $= 0.5$ mm, ar2 $= 1.5$ mm,

$$(Y_{mn}) = \begin{pmatrix} 13 & 2+i5 & -1.5-i2 & 1-i2 & -0.7+i1.3 & 0.5+i & -0.3-i0.5 & 0.1-i0.2 \\ 2+i5 & 29 & 2+i4 & -1.5-i3 & 1-i2 & -0.7+i1.3 & 0.5+i & -0.3-i0.5 \\ -1.5-i2 & 2+i4 & 60 & 2+i4 & -1.5-i3 & 1-i2 & -0.7+i1.3 & 0.5+i \\ 1-i2 & -1.5-i3 & 2+i4 & 85 & 2+i4 & -1.5-i3 & 1-i2 & -0.7+i1.3 \\ -0.7+i1.3 & 1-i2 & -1.5-i3 & 2+i4 & 85 & 2+i4 & -1.5-i3 & 1-i2 \\ 0.5+i & -0.7+i1.3 & 1-i2 & -1.5-i3 & 2+i4 & 60 & 2+i4 & -1.5-i2 \\ -0.3-i0.5 & 0.5+i & -0.7+i1.3 & 1-i2 & -1.5-i3 & 2+i4 & 29 & 2+i5 \\ 0.1-i0.2 & -0.3-i0.5 & 0.5+i & -0.7+i1.3 & 1-i2 & -1.5-i2 & 2+i5 & 13 \end{pmatrix}$$

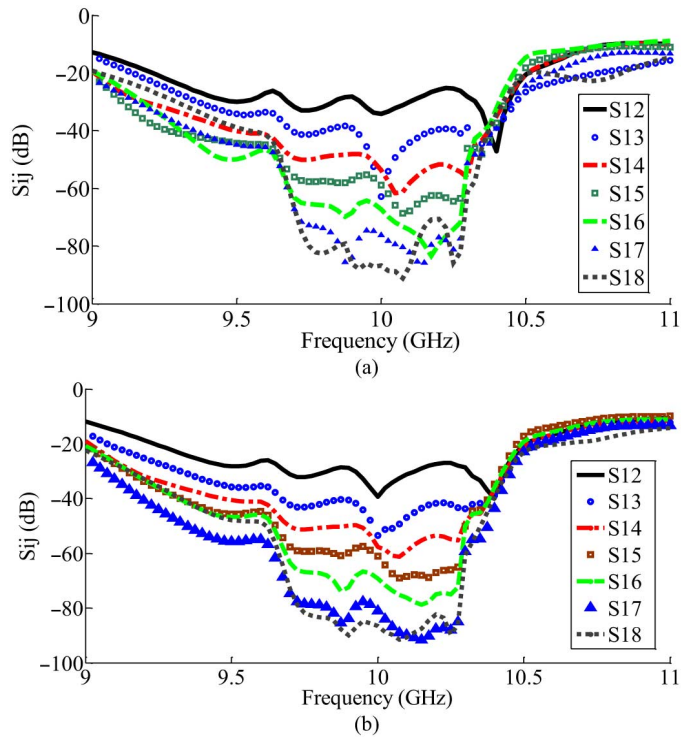

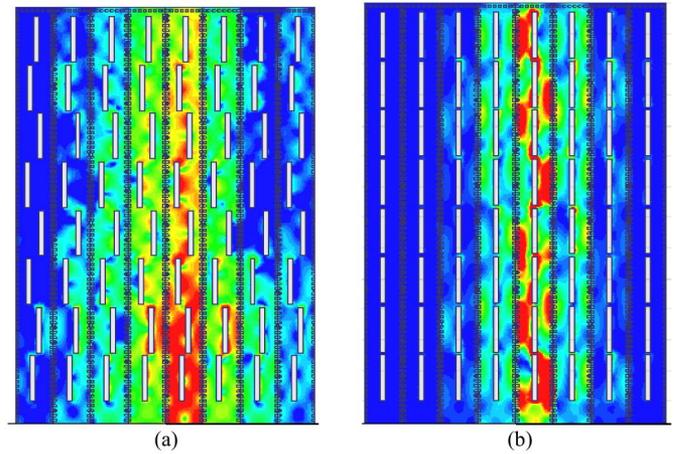

Fig. 14. Isolation of any ports for the proposed antenna. (a) Simulated. (b) Measured.

Fig. 15. Magnitude of the electric current on the upper layer. (a) Conventional slotted array antenna (with slots distanced from the centerline). (b) Proposed ridged slotted array antenna (with slots located on the centerline of the upper layer).

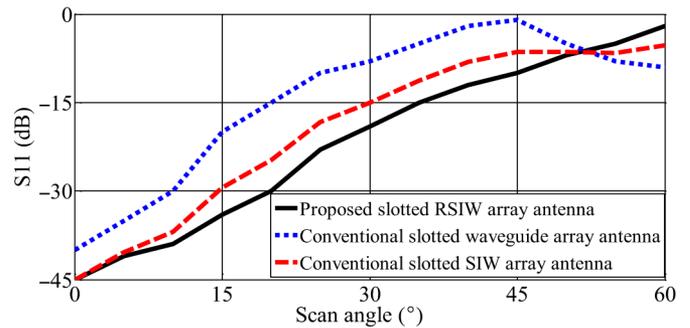

Fig. 16. S11 versus different scan angles in the radiation patterns of the conventional slotted waveguide and SIW array antennas and the proposed slotted RSIW array antenna.

and LS = 14.5 mm and the distance between the linear arrays in the E-plane equals 13 mm. Fig. 14 shows the measured and simulated $S_{ij}$ for the proposed slotted RSIW array antenna. As shown in Fig. 14, there is an isolation of at least 20 dB between different ports.

## IV. PLANAR ARRAY

However, the width of SIW structures is less than $\lambda_0/2$ and have a better performance for scanning the E-plane in comparison with the waveguide structures, SIW structures face limitations in scanning operations due to their low heights and the presence of dielectric in them and consequently high coupling between the elements.

The magnitude of the electric current on the upper layer of the proposed ridged slotted array antenna, with slots located on the centerline of the structure, and the conventional slotted array antenna, with slots distanced from the centerline, is presented in Fig. 15. In these arrays, port 4 is excited and all other input ports are matched. The magnitude of the electric current on the linear arrays adjacent to the excited linear array in the proposed ridged slotted array antenna is at a lower level compared to that of the conventional array antenna. This results in a lower mutual coupling between the excited port and all other ports.

Fig. 14 contains isolation values between ports of the eight-element array antenna. The ports possess good isolation characteristics and also a good agreement between simulation and measurement results is observed. In fact, this good isolation property causes the planar array to have low mutual coupling values. Therefore, the antenna array design can be implemented easily and with higher accuracy. Also, this low mutual coupling results in good scanning capabilities in phased array applications.

Fig. 16 shows the variations in S11 versus different scan angles in the radiation pattern. The maximum scan angle for a conventional slotted waveguide array antenna is about ±25° [19], while the maximum scan angle for the conventional slotted SIW array antenna is about ±40° and this value is increased to about ±45° in the proposed structure. In this situation, a good matching is observed and S11 has a value of less than −10 dB.

Fig. 16 shows the effect of mutual coupling reduction on the radiation pattern scanning behavior [33], [34]. The presented figure also shows that the proposed structure has a better performance compared to the conventional slotted SIW antenna and the conventional waveguide antenna.

In the design of the planar array antenna, Elliot's method is used again for obtaining slots' resonance lengths and proper widths of the ridges, while considering internal and external mutual coupling effects. A radiation pattern with a nonuniform distribution in the H-plane, uniform distribution in the E-plane and also an SLL of −25 dB, is the design goal. A 1–8 power divider is used for feeding the planar array, as shown in [31], which functions as a microstrip to SIW transition feed. There is a tapering from the 50-$\Omega$ microstrip impedance to the input impedance of the SIW structure.

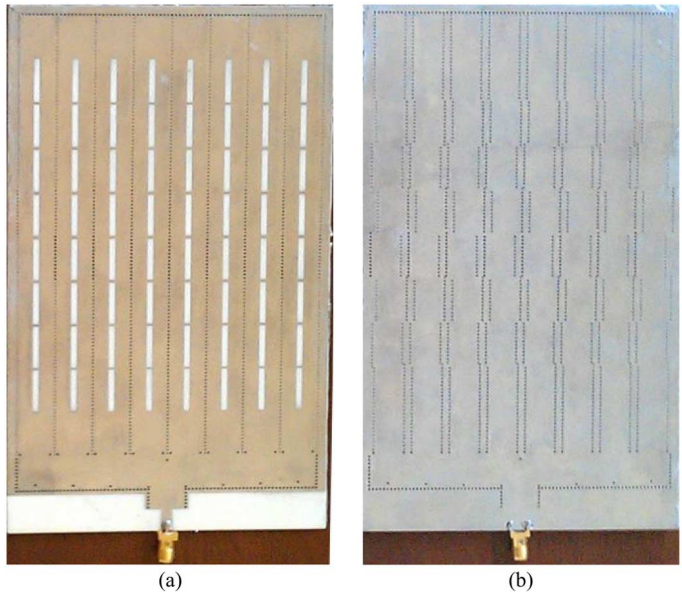

Fig. 17. Fabricated proposed planar array antenna. (a) Top plane. (b) Bottom plane.

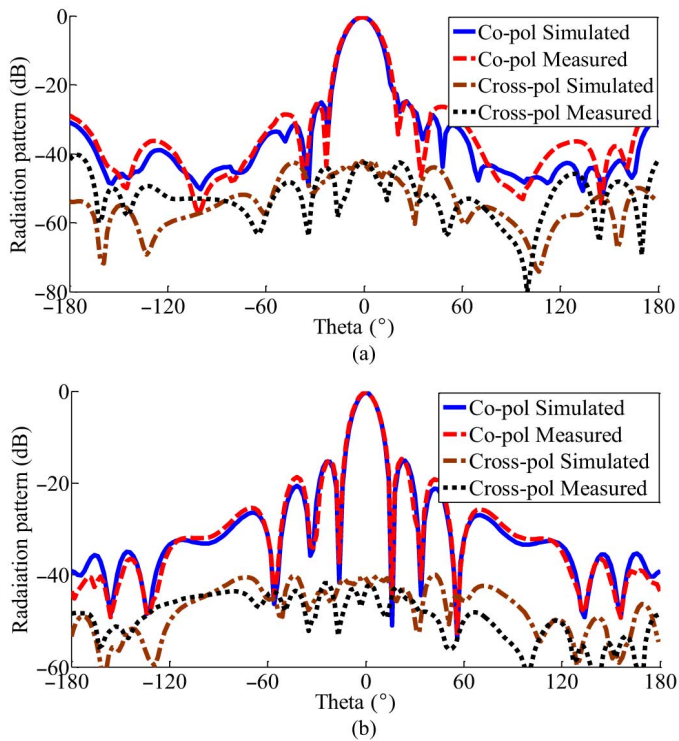

Fig. 18. Measured and simulated co- and cross-polarization radiation patterns at 10 GHz. (a) H-plane. (b) E-plane.

The $8 \times 8$ fabricated planar array antenna is depicted in Fig. 17. The proposed slotted RSIW planar array antenna is fabricated on a dielectric substrate with a thickness of 0.8 mm, a dielectric constant of 3.3, and a loss tangent of 0.002 at the frequency of 10 GHz. The fabricated antenna has dimensions of $175 \times 110$ mm$^2$.

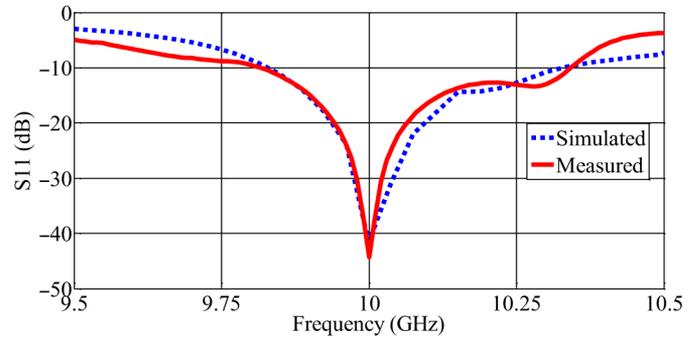

Fig. 19. Measured and simulated return loss values of the proposed planar array antenna.

The measured and simulated co-polarization and cross-polarization values of the E-plane and H-plane radiation patterns of the $8 \times 8$ array antenna at 10 GHz are presented in Fig. 18. The proposed structure has an SLL of $-25$ dB and $-15$ dB in the E-plane and H-plane, respectively. There is a good agreement between measurement and simulation results. Measurement and simulation gains of the proposed antenna are about 20 dBi and radiation efficiency is 94% in the presence of dielectric in the simulated structure; also, the measured radiation efficiency value is about 89%.

The measured and simulated return loss values are shown in Fig. 19. Regarding the measurement results, the 10-dB return loss bandwidth for the proposed planar array antenna consists of the 9.7- to 10.3-GHz frequency band. Also, a good agreement between simulation and measurement results is observed.

## V. Conclusion

A collinear longitudinal slotted RSIW array antenna is presented. Placing two ridges with different widths close to the narrow walls of the SIW causes the slots located on the centerline to radiate and also improves cross polarization of the proposed structure, compared to that of the conventional structures. The basic principles of linear antenna array design, with internal and external mutual coupling effects included, are presented. An SLL of $-25$ dB and an impedance bandwidth of 6% are achieved. Also, a planar array antenna, composed of the proposed linear array antenna, is designed and fabricated. Measurement results show good agreement with simulation results. Isolation between linear array ports is at a high level. The planar array antenna has a cross polarization of $-43$ dB and a gain of 20 dBi.

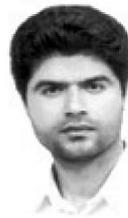

**Alireza Mallahzadeh** (M'12–SM'15) received the B.S. degree in electrical engineering from Isfahan University of Technology, Isfahan, Iran, in 1999, and the M.S. and Ph.D. degrees in electrical engineering from Iran University of Science and Technology, Tehran, Iran, in 2001 and 2006, respectively. He is a Member of Academic Staff, Faculty of Engineering, Shahed University, Tehran, Iran.

He has participated in many projects relative to antenna design. His research interests include numerical modeling and microwaves.

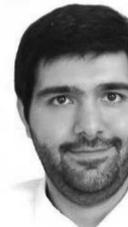

**Sajad Mohammad-Ali-Nezhad** (M'15) received the B.Sc. degree in electronic engineering from Shahid Chamran University, Ahwaz, Iran, in 2008, and the M.Sc. and Ph.D. degrees in communication engineering from Shahed University, Tehran, Iran, in 2010 and 2015, respectively.

Currently, he is the Head of the Department of Electrical and Electronic Engineering, University of Qom, Qom, Iran. His research interests include leaky-wave antennas, printed circuit antennas, array antennas, phased array antennas, MIMO antennas, RFID tag antennas, frequency-selective surface, electromagnetic compatibility, microwave filters, and electromagnetic theory.